\documentclass[journal]{IEEEtran}
\ifCLASSINFOpdf
\else
\fi
\usepackage{cite}
\usepackage{amsmath}
\usepackage[pdftex]{graphicx} 
\usepackage{subfig}
\usepackage{array} 
\usepackage{fixltx2e}

\usepackage{booktabs} 
\usepackage{amssymb}
\usepackage{mathtools}
\usepackage{cancel} 
\usepackage{bm} 
\usepackage{esint} 
\usepackage{amsthm}

\usepackage{tikz}
\usepackage{tikz-3dplot}
\usetikzlibrary{backgrounds,3d,shadings,shapes.misc,decorations.pathmorphing,shapes,calc,fadings,shadows.blur}

\usepackage{amssymb,amsmath}

\usepackage{bm}
\newcommand{\bs}[1]{\bm{\mathrm{#1}}}

\newcommand{\vect}[1]{\vec{#1}}
\newcommand{\nhat}{\ensuremath{\hat{n}}}

\newcommand{\vecprime}[1]{\vect{#1}^{\,\prime}}

\renewcommand{\r}{\left(\vect{r}\right)}
\newcommand{\rp}{\left(\vecprime{r}\right)}

\newcommand{\rrpki}[1][]{\left(k_i, \vect{r}, \vecprime{r}\right)}

\newcommand{\matr}[1]{\bs{#1}}

\newcommand{\junk}[1] {}

\def\XXint#1#2#3{{\setbox0=\hbox{$#1{#2#3}{\int}$}
\vcenter{\hbox{$#2#3$}}\kern-.5\wd0}}

\newcommand*\widebar[1]{%
  \hbox{%
    \vbox{%
      \hrule height 0.5pt 
      \kern0.3ex
      \hbox{%
        \kern-0.05em
        \ensuremath{#1}%
        \kern-0.05em
      }%
    }%
  }%
} 

\renewcommand{\epsilon}{\varepsilon}

\newcommand{\opL}{\ensuremath{\mathcal{L}}} 
\newcommand{\opK}{\ensuremath{\mathcal{K}}} 

\newcommand{\surf}{\mathcal{S}}

\newcommand{\Lmat}[1][]{{\matr{L}_{#1}}}
\newcommand{\LmatA}[1][]{{\matr{L}^{(A)}_{#1}}}
\newcommand{\LmatPhi}[1][]{{\matr{L}^{(\phi)}_{#1}}}
\newcommand{\Kmat}[1][]{{\matr{K}_{#1}}}

\newcommand{\Pxout}{\ensuremath{\matr{I}_{\times}}} 

\newcommand{\Pxin}{\ensuremath{\matr{I}_{\times}}} 

\newcommand{\figref}[1]{Fig.~\ref{#1}}
\newcommand{\secref}[1]{Section~\ref{#1}}

\usepackage{tcolorbox}

\usepackage{textpos}

\newcommand{\mySubtitle}[1]%
{%
	\begin{textblock}{14.0}(0.7, 2.9)
		\textbf{#1}%
	\end{textblock}%
}%

\newcommand{\redcol}{black}

\newcommand{\red}[1]{\textcolor{\redcol}{#1}}
\newcommand{\green}[1]{\textcolor{green!0!black}{#1}}

\newcommand{\Ecolor}[1]{{#1}}
\newcommand{\Hcolor}[1]{{#1}}
\newcommand{\Jcolor}[1]{{#1}}

\newcommand{\Er}[1][]{\Ecolor{\vect{E}_{#1}\r}}

\newcommand{\Etr}[1][]{\Ecolor{\nhat_{#1} \times \vect{E}_{#1}\r}}
\newcommand{\Etrp}[1][]{\Ecolor{\nhat_{#1}' \times \vect{E}_{#1}\rp}}

\newcommand{\Emat}[1][]{\Ecolor{\matr{E}_{#1}}}

\newcommand{\Hr}[1][]{\Hcolor{\vect{H}_{#1}\r}}

\newcommand{\Htr}[1][]{\Hcolor{\nhat_{#1} \times \vect{H}_{#1}\r}}
\newcommand{\Htrp}[1][]{\Hcolor{\nhat_{#1}' \times \vect{H}_{#1}\rp}}
\newcommand{\Hmat}[1][]{\Hcolor{\matr{H}_{#1}}}

\newcommand{\Jmat}[1][]{\ensuremath{\Jcolor{\matr{J}_{#1}}}}

\newcommand{\Grrpki}[1][]{\ensuremath{\green{G\rrpki}}}

\newcommand{\rhomat}[1][]{\red{\matr{\rho}_{#1}}}

\makeatletter
\newlength\numerator@height
\newlength\frac@height
\newsavebox\numerator@box
\newsavebox\frac@box

\newcommand\dfracparens[3]{%
	\sbox{\numerator@box}{\ensuremath{#1}}%
	\sbox{\frac@box}{\ensuremath{\dfrac{#1}{#2}}}%
	\settoheight{\frac@height}{\usebox{\frac@box}}%
	\settoheight{\numerator@height}{\usebox{\numerator@box}}%
	\addtolength{\frac@height}{-\numerator@height}%
	\usebox{\frac@box}%
	\raisebox{\frac@height}{%
		\( \left( {#3} \right)
		\)}%
}
\makeatother

\newcommand{\revcolor}[1]{\textcolor{black}{#1}}
\newcommand{\shortcolor}[1]{\textcolor{black}{#1}}

\begin{document}
%
\title{SLIM: A Well-Conditioned~Single-Source Boundary~Element Method for Modeling Lossy~Conductors in Layered~Media}
%
%
%

\author{Shashwat~Sharma,~\IEEEmembership{Student Member,~IEEE,}
        and~Piero~Triverio,~\IEEEmembership{Senior Member,~IEEE}
\thanks{Accepted to the IEEE Antennas and Wireless Propagation Letters (Volume: 19, Issue: 12, Dec. 2020, DOI: 10.1109/LAWP.2020.3022593).}
\thanks{S. Sharma and P. Triverio are with the Edward S. Rogers Sr. Department of Electrical \& Computer Engineering, University of Toronto, Toronto,
ON, M5S 3G4 Canada, e-mails: shash.sharma@mail.utoronto.ca, piero.triverio@utoronto.ca.}
\thanks{This work was supported by the Natural Sciences and Engineering Research 
	Council of Canada (Collaborative Research and Development Grants 
	program), by Advanced Micro Devices, and by CMC Microsystems.}}

%
%

\markboth{IEEE Antennas and Wireless Propagation Letters}%
{Sharma \MakeLowercase{\textit{et al.}}: Bare Demo of IEEEtran.cls for IEEE Journals}
%



\maketitle

\begin{abstract}
  The boundary element method (BEM) enables the efficient electromagnetic modeling of lossy conductors with a surface-based discretization.
  Existing BEM techniques for conductor modeling require either expensive dual basis functions or the use of both single- and double-layer potential operators to obtain a well-conditioned system matrix.
  The computational cost is particularly significant when conductors are embedded in stratified media, and the expensive multilayer Green's function (MGF) must be used.
  In this work, a novel single-source BEM formulation is proposed, which leads to a well-conditioned system matrix without the need for dual basis functions.
  The proposed single-layer impedance matrix~(SLIM) formulation does not require the double-layer potential to model the background medium, which reduces the cost associated with the MGF.
  The accuracy and efficiency of the proposed method are demonstrated through realistic examples drawn from different applications.
\end{abstract}


%
\IEEEpeerreviewmaketitle

\section{Introduction}

\IEEEPARstart{T}{he} need for full-wave electromagnetic (EM) modeling of lossy conductors arises in many applications.
The design of high-speed on-chip interconnects requires an accurate prediction of signal propagation and cross-talk, which is heavily influenced by the frequency-dependent variation of skin depth~\cite{MTL_Paul}.
Likewise, the quantification of losses in the sub-wavelength unit cells of metasurfaces and metamaterials requires modeling conductors as imperfect~\cite{SRRmetaCond}.

Electromagnetic modeling of lossy conductors with the finite element method~\cite{FEMJin} or volume integral equations~\cite{VolIE01,PEEC01,fastmaxwell} requires an extremely fine volumetric discretization of the structure to capture skin effect at high frequencies.
In contrast, the boundary element method (BEM), which is based on a surface integral representation of Maxwell's equations~\cite{gibson}, requires meshing only on the interfaces between objects.
This reduction of dimensionality allows capturing the skin effect without the need for an extremely fine mesh,
making the BEM an appealing approach for modeling penetrable media.

Conductor modeling with the BEM requires describing an interior problem to capture the physics within objects, and an exterior problem to model coupling between them.
Approximate techniques for the interior problem such as the Leontovich surface impedance boundary condition (SIBC)~\cite{SIBC} are not accurate near corners or at low frequencies, where skin effect has not yet developed.
The generalized impedance boundary condition (GIBC)~\cite{GIBC} and related formulations~\cite{agibc,gibcHmatDanJiao} are well conditioned and applicable over a broad frequency range, but require both single- and double-layer potential operators~\cite{book:colton} for modeling the exterior problem.
When conductors are embedded in stratified background media, the multilayer Green's function (MGF)~\cite{MGF01} is required for computing the single-layer potential, and its curl is required for the double-layer potential~\cite{MGF01}.
Both the MGF and its curl are several times more expensive to compute than the homogeneous Green's function.
Therefore, having to compute both the single- and double-layer operators for layered media has a significant cost, and increases code complexity.
Other methods such as the enhanced augmented electric field integral equation (eAEFIE)~\cite{eaefie02} require not only the same operators as the GIBC, but also the use of expensive dual basis functions~\cite{BCorig} to obtain a well-conditioned system matrix~\cite{TAP2020}.

The differential surface admittance~(DSA) approach~\cite{DSA01,DSA08,UTK_AWPL2017,EPEPS2017} is a single-source formulation which
requires only the single-layer potential operator of the exterior problem.
However, the DSA formulation also leads to an ill-conditioned system matrix unless dual basis functions are employed~\cite{APS2019,DSA_Calderon_HDC}.
A single-source impedance-based~(SSI) formulation was proposed as a remedy~\cite{APS2020}, but the resulting system matrix is still not as well conditioned as the GIBC.

In this work, a novel single-source formulation is proposed, which offers the advantages of both the GIBC and the DSA, while overcoming their shortcomings.
Like the DSA, the proposed single-layer impedance matrix (SLIM) formulation involves only a differential electric current density as the source~\cite{DSA01}.
Therefore, the SLIM approach does not require the double-layer potential operator for the exterior problem, unlike the GIBC, which leads to a smaller matrix fill time.
Furthermore, the SLIM formulation leads to a well-conditioned system matrix without the need for dual basis functions, unlike the DSA or SSI approaches.


\section{Proposed Formulation}\label{sec:prop:slim}

In the following, time-harmonic fields with a time dependence of $e^{j\omega t}$ are assumed, where $\omega$ is the cyclical frequency and $j = \sqrt{-1}$.
Primed coordinates represent source points, while unprimed coordinates represent observation points.
We first consider a structure composed of a single object occupying volume $\mathcal{V}$, bounded by a surface $\surf$ with outward unit normal vector $\nhat$.
The proposed method will be generalized to structures with multiple objects in \secref{sec:prop:slim:multiobject}.
The object is assumed to be homogeneous with permittivity $\epsilon$, permeability $\mu$, and electrical conductivity $\sigma$.
The object may be embedded in a layered medium, where the $l^{\mathrm{th}}$ layer is denoted by $\mathcal{V}_l$, has permittivity $\epsilon_l$ and permeability $\mu_l$.
We define $\mathcal{V}_0 = \bigcup_{l=1}^{N_l}\mathcal{V}_l$.
The system may be excited with an incident plane wave, $[\Er[\mathrm{inc}], \Hr[\mathrm{inc}]]$, $\vect{r} \in \mathcal{V}_0$, or through lumped ports~\cite{gope}, resulting in a field distribution $[\Er, \Hr]$ for $\vect{r} \in \mathcal{V}$.

\subsection{Interior Problem}\label{sec:prop:slim:interior}

\subsubsection{Original Configuration}\label{sec:prop:slim:interior:orig}
The tangential electric and magnetic fields on $\surf$ can be related via the magnetic field integral equation~(MFIE)~\cite{gibson},
\begin{multline}
  j\omega\epsilon\, \nhat \times \opL \left[\Etrp\right]\r
  + \nhat \times \opK \left[\Htrp\right]\r\\
  - \dfrac{1}{2} \Htr = 0\label{eq:MFIE1}
\end{multline}
for $\vect{r}, \vect{r}\,' \in \surf$.
Using the MFIE \eqref{eq:MFIE1} rather than the electric field integral equation (EFIE) will lead to better conditioning of the final system matrix.
Definitions of the integro-differential operators $\opL$ and $\opK$ may be found in literature~\cite{gibson}, and involve the homogeneous Green's function of the object's material.

A triangular mesh is generated for the surface of the object.
Equation~\eqref{eq:MFIE1} is discretized by expanding $\Etr$ and $\Htr$ with $\mathrm{RWG}$ basis functions~\cite{RWG} and tested with rotated $\nhat \times \mathrm{RWG}$ functions.
This leads to the matrix equation
\begin{align}
  -j\omega\epsilon\, \Lmat \Emat - \left( \Kmat - \dfrac{1}{2}\Pxin \right) \Hmat &= \matr{0},\label{eq:MFIE1dis}
\end{align}
where $\matr{L}$ and $\matr{K}$ are the discretized $\opL$ and $\opK$ operators.
For highly conductive media, specialized integration routines must be used for assembling $\matr{L}$ and $\matr{K}$, to capture the fast oscillations of the Green's function~\cite{GIBC,eaefie02}.
Matrix $\Pxout$ is the identity operator obtained by testing $\mathrm{RWG}$ with $\nhat\times\mathrm{RWG}$ functions.
Column vectors $\Emat$ and $\Hmat$ contain the coefficients of the basis functions associated with $\Etr$ and $\Htr$, respectively.

\subsubsection{Equivalent Configuration}\label{sec:prop:slim:interior:eq}

Next, we use the surface equivalence principle~\cite{EMharrington} to replace the conductive object with the background material in which it resides, while requiring that $\Etr$ remains unchanged for $\vect{r}\in\surf$~\cite{DSA01}.
An equivalent electric current density, $\vect{J}_\Delta$,
\begin{align}
  \vect{J}_{\Delta}\r &= \Htr - \Hcolor{\nhat\, \times\, }\Hr[\mathrm{eq}],\label{eq:Jdelta}
\end{align}
must be introduced on $\surf$ to keep fields in $\mathcal{V}_0$ unchanged.
In \eqref{eq:Jdelta}, $\Hcolor{\nhat\,\times}\Hr[\mathrm{eq}]$ is the tangential magnetic field on $\surf$ in the equivalent configuration.
Enforcing an unchanged tangential electric field will later avoid the need for the double-layer potential operator of the exterior problem~\cite{DSA01}.

Vector fields $\vect{J}_{\Delta}$, $\Htr$ and $\Hcolor{\nhat\, \times\, }\Hr[\mathrm{eq}]$ are now expanded with $\mathrm{RWG}$ basis functions to obtain the discrete version of \eqref{eq:Jdelta},
\begin{align}
  \Jmat[\Delta] = \Hmat - \Hmat[\mathrm{eq}],\label{eq:Jdelta2}
\end{align}
where column vectors $\Jmat[\Delta]$ and $\Hmat[\mathrm{eq}]$ contain the coefficients of the basis functions associated with $\vect{J}_{\Delta}$ and $\Hcolor{\nhat\, \times\, }\Hr[\mathrm{eq}]$, respectively.

In the equivalent configuration, the $\Etr$ and $\Hcolor{\nhat\, \times\, }\Hr[\mathrm{eq}]$ can also be related via the EFIE~\cite{gibson},
which after discretization reads
\begin{align}
  j\omega\mu_l\, \Lmat[l] \Hmat[\mathrm{eq}] - \left( \Kmat[l] - \dfrac{1}{2}\Pxout \right) \Emat &= \matr{0},\label{eq:EFIE1dis}
\end{align}
where $\Lmat[l]$ and $\Kmat[l]$ are the discretized $\opL$ and $\opK$ operators involving the homogeneous Green's function associated to $\mathcal{V}_l$.

\subsection{Exterior Problem}\label{sec:prop:slim:exterior}

To enable broadband simulations of multiscale objects, the augmented EFIE (AEFIE)~\cite{aefie2} is employed to model the exterior problem.
The charge density $\rho_\Delta$ associated to $\vect{J}_{\Delta}$ is introduced as an additional unknown, and discretized with pulse basis functions~\cite{aefie2}.
The total electric field tangential to $\surf$ is written in discretized form as
\begin{align}
  jk_0\LmatA[\mathrm{m}] \Jmat[\Delta] + c_0\matr{D}^T\LmatPhi[\mathrm{m}]\matr{B}\rhomat[\Delta] + \eta_0^{-1}\Pxout\Emat = \eta_0^{-1}\Emat[\mathrm{inc}],\label{eq:AEFIEextdis}
\end{align}
where $\LmatA[\mathrm{m}]$ and $\LmatPhi[\mathrm{m}]$ are the discretized vector and scalar potential parts of the $\opL$ operator, respectively, and $\rhomat[\Delta]$ contains coefficients of $\rho_\Delta$.
Subscript $\mathrm{m}$ indicates that the multilayer Green's function of the stratified background medium is used.
Column vector $\Emat[\mathrm{inc}]$ is related to the incident electric field.
Quantities $k_0$, $\eta_0$, and $c_0$ are, respectively, the wave number, wave impedance, and speed of light in free space.
The differential current and charge densities are also related via the continuity equation,
\begin{align}
  \matr{F}\matr{D}\Jmat[\Delta] + jk_0c_0\matr{I}\rhomat[\Delta] = \matr{0},\label{eq:contdis}
\end{align}
where $\matr{I}$ is the identity matrix. Definitions of the sparse matrices $\matr{D}$, $\matr{B}$ and $\matr{F}$ may be found in~\cite{aefie2}.

\subsection{Final System Matrix}\label{sec:prop:slim:system}\label{sec:prop:slim:multiobject}

The goal now is to utilize equations \eqref{eq:MFIE1dis} and \eqref{eq:Jdelta2}--\eqref{eq:contdis} to derive a well-conditioned final system matrix.
First, the tangential electric field $\Emat$ is expressed in terms of the tangential magnetic field $\Hmat$ with a rearrangement of \eqref{eq:MFIE1dis},
\begin{align}
  \Emat = \underbrace{\dfrac{-1}{j\omega\epsilon} \left(\Lmat\right)^{-1} \left(\Kmat - \dfrac{1}{2}\Pxout\right)}_{\triangleq\matr{Z}}\Hmat.\label{eq:MFIEinZin}
\end{align}
Next, the equivalent tangential magnetic field $\Hmat[\mathrm{eq}]$ is expressed in terms of $\Emat$ with a rearrangement of \eqref{eq:EFIE1dis},
\begin{align}
  \Hmat[\mathrm{eq}] = \underbrace{\dfrac{1}{j\omega\mu_{l}} \left(\matr{L}_{\mathrm{eq}}\right)^{-1} \left(\matr{K}_{\mathrm{eq}} - \dfrac{1}{2}\matr{I}_{\mathrm{x}}\right)}_{\triangleq\matr{Y}_{\mathrm{eq}}} \Emat.\label{eq:EFIEeqYeq}
\end{align}
Matrices $\matr{Z}$ and $\matr{Y}_{\mathrm{eq}}$ may be interpreted, respectively, as the surface impedance of the original object~\cite{GIBC} and the surface admittance of the object replaced with the surrounding medium~\cite{UTK_AWPL2017}.

Using \eqref{eq:MFIEinZin} in \eqref{eq:EFIEeqYeq}, we may express $\Hmat[\mathrm{eq}]$ in terms of $\Hmat$ as
\begin{align}
  \Hmat[\mathrm{eq}] = \matr{Y}_{\mathrm{eq}} \matr{Z} \Hmat.\label{eq:EFIEeqYZ}
\end{align}
Equations \eqref{eq:Jdelta2}, \eqref{eq:MFIEinZin} and \eqref{eq:EFIEeqYZ} now allow us to write \eqref{eq:AEFIEextdis} and \eqref{eq:contdis} in terms of only $\Hmat$ and $\rhomat[\Delta]$, to obtain the final system
\begin{align}
  \begin{bmatrix}
    jk_0 \LmatA[\mathrm{m}] + \matr{C} &
    \matr{D}^T\LmatPhi[\mathrm{m}]\matr{B} \\
    \matr{F}\matr{D}\left(\matr{I} - \matr{Y}_{\mathrm{eq}}\matr{Z}\right) & jk_0\matr{I}
  \end{bmatrix}
  \begin{bmatrix}
    \Hmat \\ c_0\rhomat[\Delta]
  \end{bmatrix} =
  \begin{bmatrix}
    \Emat[\mathrm{inc}]/\eta_0 \\ \matr{0}
  \end{bmatrix},\label{eq:AEFIEoutJ}
\end{align}
where
\begin{align}
  \matr{C} = \left(-jk_0 \LmatA[\mathrm{m}] \matr{Y}_{\mathrm{eq}} + \eta_0^{-1}\Pxout\right)\matr{Z}.\label{eq:Cdef}
\end{align}
Equation~\eqref{eq:AEFIEoutJ} is the proposed SLIM formulation.
The top left block of~\eqref{eq:AEFIEoutJ} has a similar form to the corresponding matrix block in the augmented GIBC formulation~\cite{agibc}, but with a different definition of the matrix $\matr{C}$, which may be interpreted as a correction to the well-conditioned vector potential operator for perfect electric conductors, $jk_0\LmatA[\mathrm{m}]$.
It has been shown that the GIBC formulation leads to a well-conditioned system matrix without the need for dual basis functions~\cite{TAP2020}, and the similarity to this form is responsible for the good conditioning of the SLIM method compared to existing single-source BEM formulations~\cite{UTK_AWPL2017,APS2020}.
\revcolor{In particular, $\LmatA[\mathrm{m}]$, which is well conditioned~\cite{KleinmanErrCond}, has a dominant contribution to the overall conditioning of the system in~\eqref{eq:AEFIEoutJ}, compared to the ``correction'' matrix~$\matr{C}$, which accounts for the finite conductivity of the object. The dominance of $\LmatA[\mathrm{m}]$ implies that even though some terms in $\matr{C}$ are not well tested, for example $\Pxout$, their impact is diminished, and the use of dual basis functions can be avoided.
The same is true for the GIBC, as confirmed in \secref{sec:results}.}

Additionally, the SLIM formulation avoids the double-layer potential operator for the exterior problem, required in the GIBC.
For objects embedded in stratified media, the single-layer potential involves the dyadic multilayer Green's function (MGF), while the double-layer potential involves its curl~\cite{MGF01}.
Avoiding the double-layer operator precludes the need to compute the curl of the MGF, which leads to a significant reduction in CPU time.
This feature also reduces code complexity, since the extraction of singular and quasistatic terms for the curl of the MGF is more challenging than for the MGF alone~\cite{qse}.

Although the SLIM formulation requires an additional matrix operator $\matr{Y}_{\mathrm{eq}}$ in the interior problem, this operator involves the homogeneous Green's function, which is cheaper to compute than the MGF.
Furthermore, for large objects, the inversions in~\eqref{eq:MFIEinZin} and~\eqref{eq:EFIEeqYeq} can be avoided by applying $\matr{Z}$ and $\matr{Y}_{\mathrm{eq}}$ iteratively with the multiple-grid adaptive integral method (AIM)~\cite{AIMbles} presented in~\cite{TAP2020}.
To efficiently solve~\eqref{eq:AEFIEoutJ} with an iterative solver, the constraint preconditioner in~\cite{aefie2} is adopted.
The matrix-vector product in~\eqref{eq:AEFIEoutJ} is accelerated with a multilayer version of the AIM~\cite{CPMT2019arxiv}.

\revcolor{%
For structures with multiple objects, matrices $\matr{Z}_i$ and $\matr{Y}_{i,\mathrm{eq}}$ are independently constructed for each object $i$.
For assembling $\matr{Z}_i$ using~\eqref{eq:MFIEinZin}, the properties and Green's function associated to the material of the $i^\mathrm{th}$ object are used.
For $\matr{Y}_{i,\mathrm{eq}}$, the properties and Green's function associated to the material surrounding the $i^\mathrm{th}$ object are used.
A block diagonal concatenation of all $\matr{Z}_i$ and $\matr{Y}_{i,\mathrm{eq}}$ provides $\matr{Z}$ and $\matr{Y}_{\mathrm{eq}}$ as
\begin{align}
	\matr{Z} &= \mathrm{diag}\begin{bmatrix}\matr{Z}_1 & \matr{Z}_2 \ldots \matr{Z}_{N_\mathrm{obj}}\end{bmatrix},\\
	\matr{Y}_\mathrm{eq} &= \mathrm{diag}\begin{bmatrix}\matr{Y}_{1,\mathrm{eq}} & \matr{Y}_{2,\mathrm{eq}} \ldots \matr{Y}_{N_\mathrm{obj},\mathrm{eq}}\end{bmatrix},
\end{align}
where $N_\mathrm{obj}$ is the number of objects.
Matrices $\matr{Z}$ and $\matr{Y}_\mathrm{eq}$ can then be used directly in~\eqref{eq:AEFIEoutJ} and~\eqref{eq:Cdef}.%
}


\section{Results}\label{sec:results}

The proposed method is compared to the GIBC~\cite{GIBC}, DSA~\cite{UTK_AWPL2017} and SSI~\cite{APS2020} formulations through numerical examples.
In each formulation, the AEFIE is used in the exterior problem~\cite{agibc,CPMT2019arxiv}, and the interior problem is accelerated with the method in~\cite{TAP2020}.
All simulations were run single-threaded on a 3\,GHz Intel Xeon CPU.
We used PETSc~\cite{petsc-web-page} for sparse matrix manipulation.
The GMRES iterative solver~\cite{gmres} with a relative residual norm of $10^{-4}$ was used for solving~\eqref{eq:AEFIEoutJ}.

\subsection{Scattering from a Copper Sphere}\label{sec:results:sphere1m}

\begin{figure}[t]
	\centering
	\includegraphics[width=\linewidth,clip=true,trim={0 0.0 0 0}]{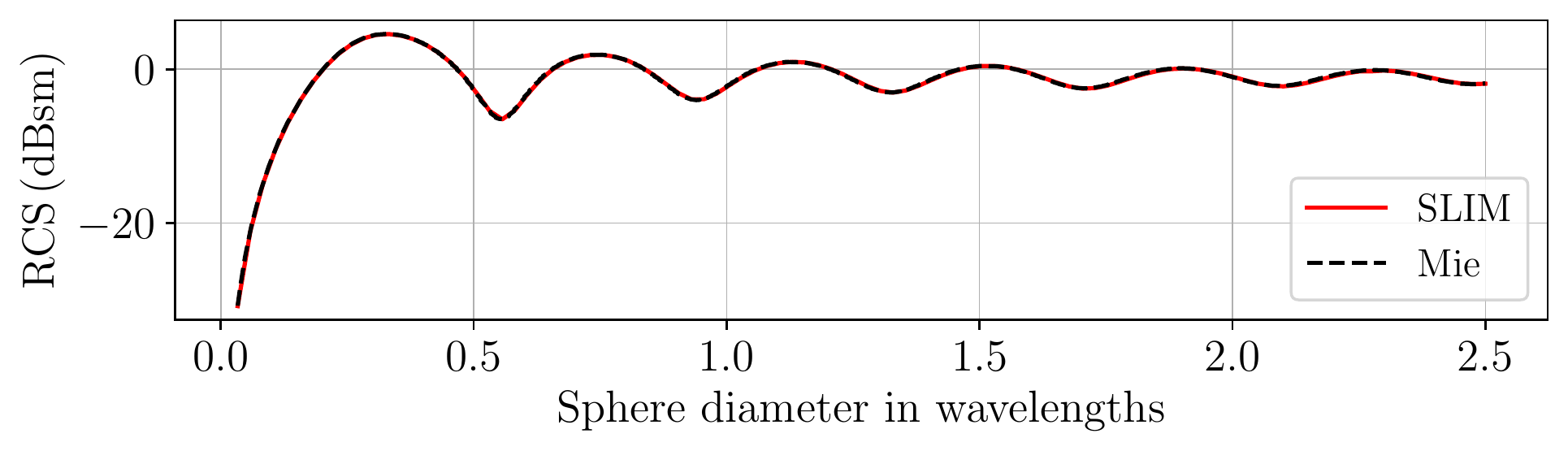}\\
  \includegraphics[width=\linewidth]{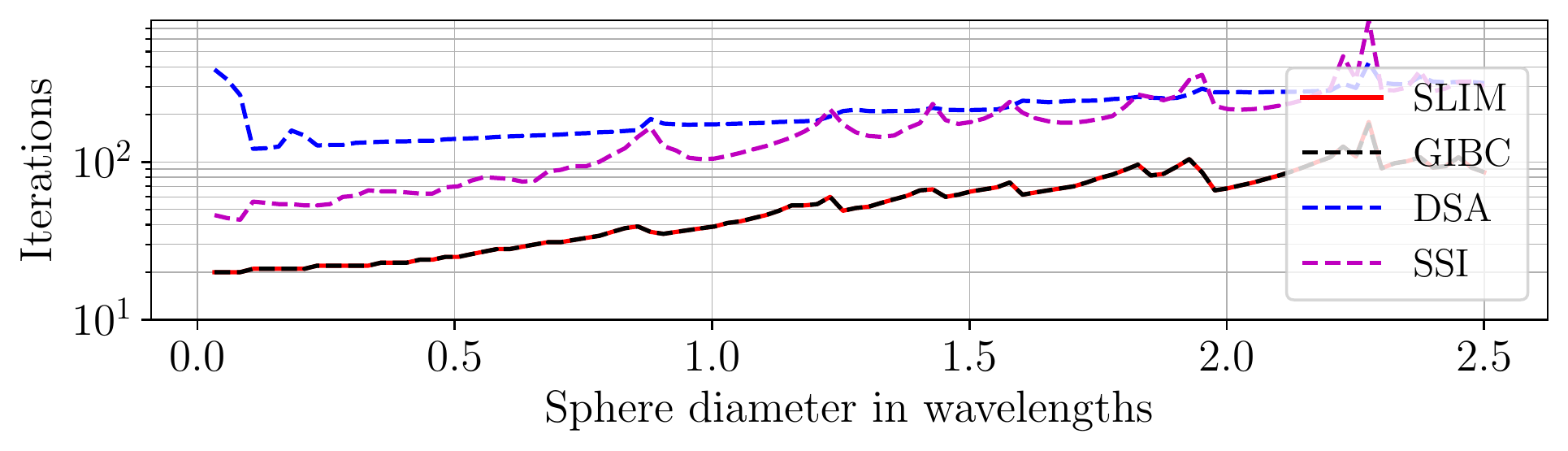}
	\caption{\shortcolor{Top panel:} monostatic RCS compared to analytical results. \shortcolor{Bottom panel:} number of iterations for the proposed, GIBC, DSA and SSI methods, for the sphere in \secref{sec:results:sphere1m}.}\label{fig:sphere1m}
\end{figure}

To validate the accuracy of the proposed method, a copper sphere ($\sigma = 5.8\times 10^7\,$S/m) with diameter $1\,$m is \revcolor{considered, meshed with $1,956$ triangles and $2,934$ edges.
The structure is}
 excited with a plane wave at frequencies ranging from $10\,$MHz to $750\,$MHz, in free space.
The monostatic radar cross section (RCS) is computed and compared with the analytical solution obtained via Mie series~\cite{gibson}.
\shortcolor{The top panel of} \figref{fig:sphere1m} indicates an excellent agreement over the entire range of frequencies considered.
\shortcolor{The bottom panel of} \figref{fig:sphere1m} shows the number of iterations required at each frequency, for the SLIM, GIBC, DSA and SSI approaches.
It is clear that the SLIM and GIBC methods lead to significantly better matrix conditioning than the DSA and SSI formulations.

\subsection{Inductor Array}\label{sec:results:indarray}

\begin{figure}[t]
  \centering
  \includegraphics[width=.90\linewidth,trim={0cm 0cm 0cm 0cm},clip=true]{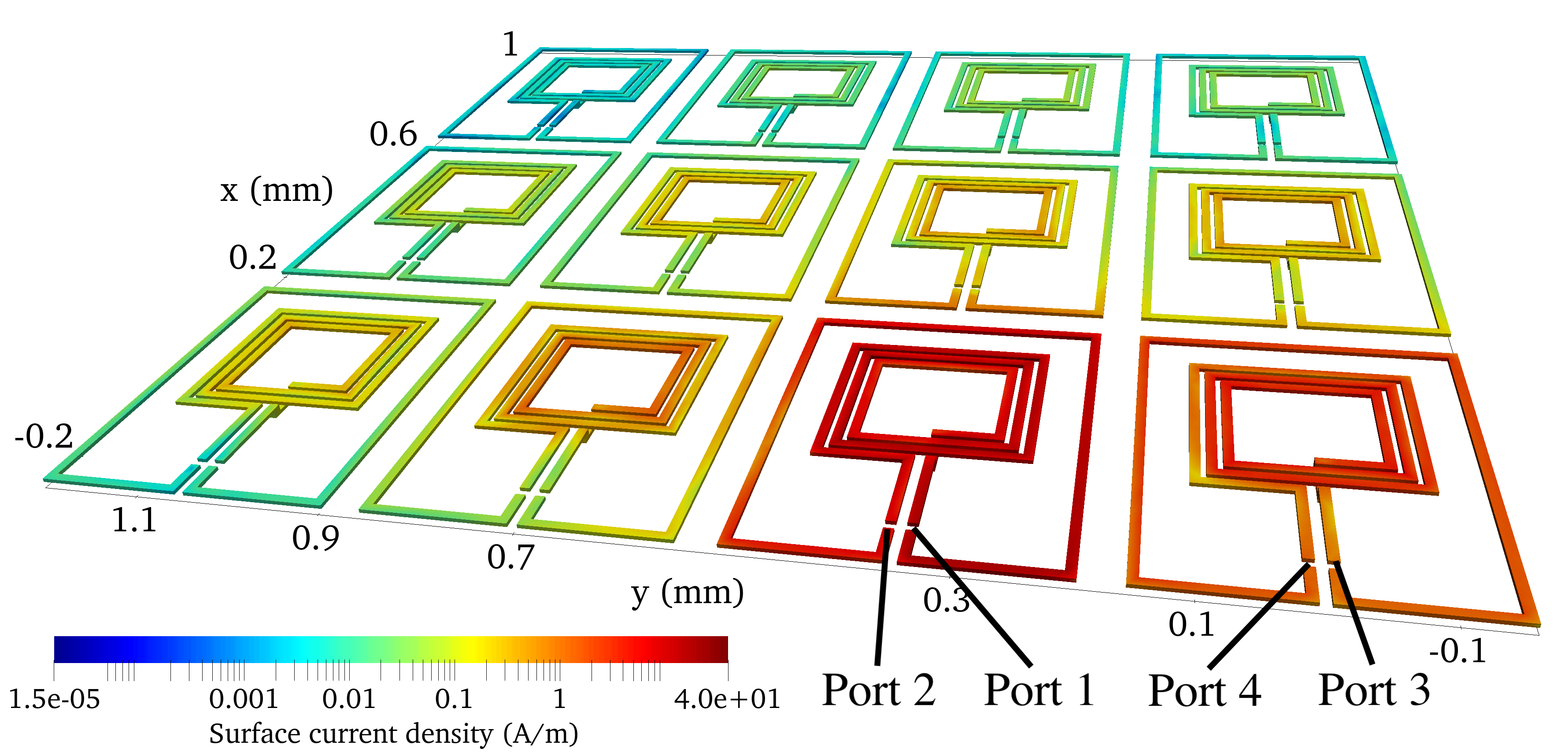}
  \caption{Geometry of the inductor array in \secref{sec:results:indarray}, with the electric surface current density at $30\,$GHz.}\label{fig:indarrayJ}
\end{figure}

\begin{figure}[t]
  \centering
  \includegraphics[width=\linewidth]{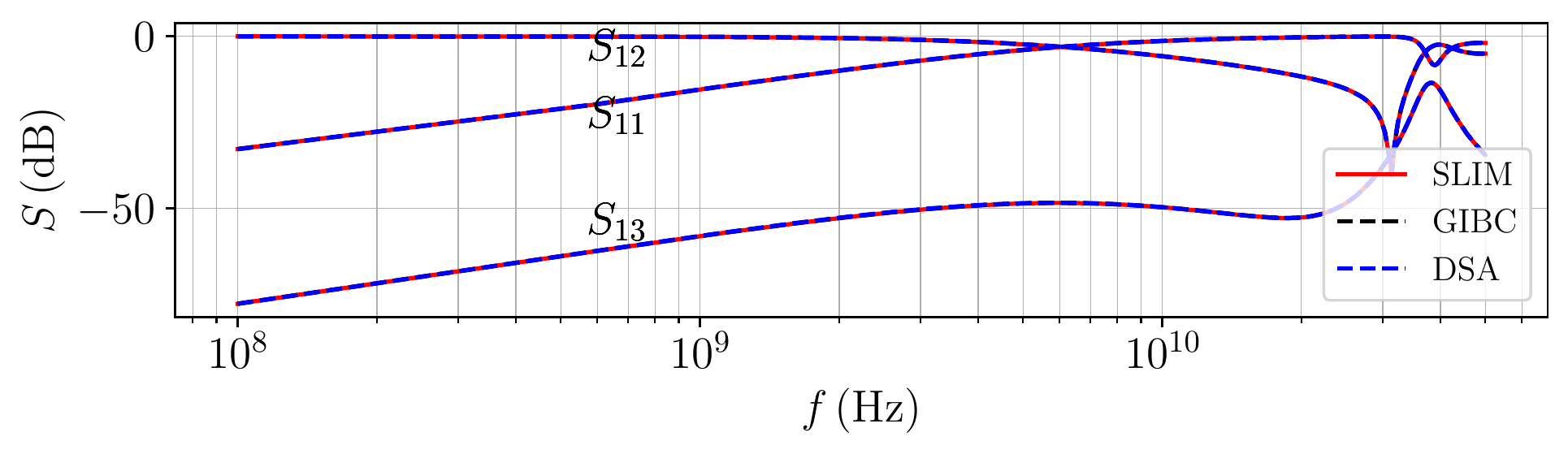}
  \caption{Comparison of $S$ parameters for for the inductor array in \secref{sec:results:indarray}.}\label{fig:indarrayS}
\end{figure}

\begin{figure}[t]
  \centering
  \includegraphics[width=\linewidth]{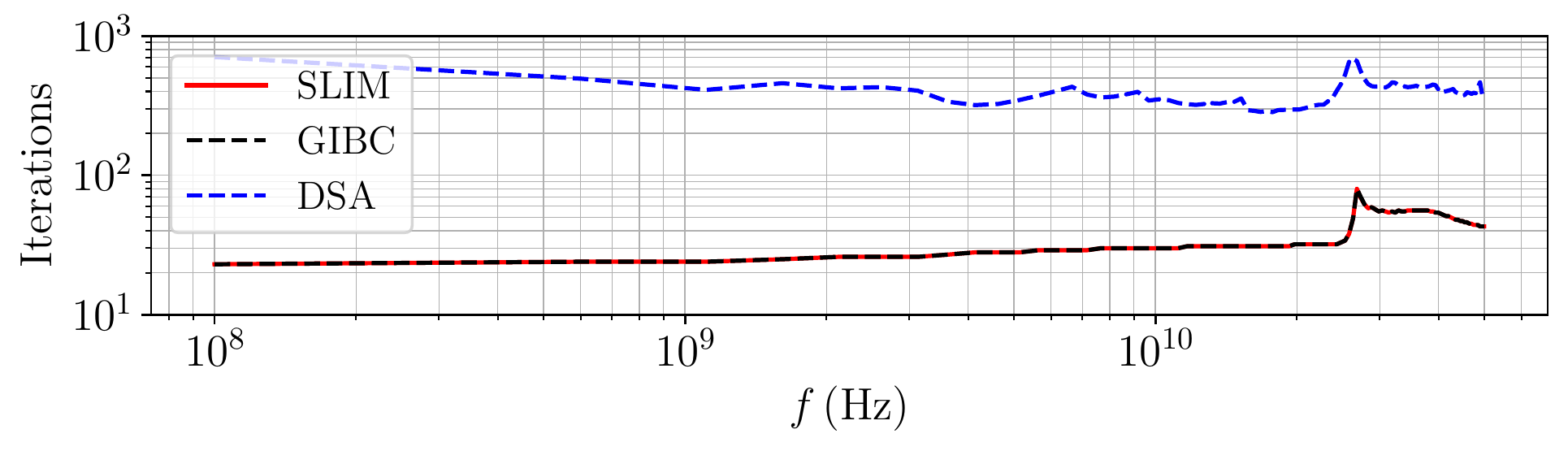}\\
  \includegraphics[width=\linewidth]{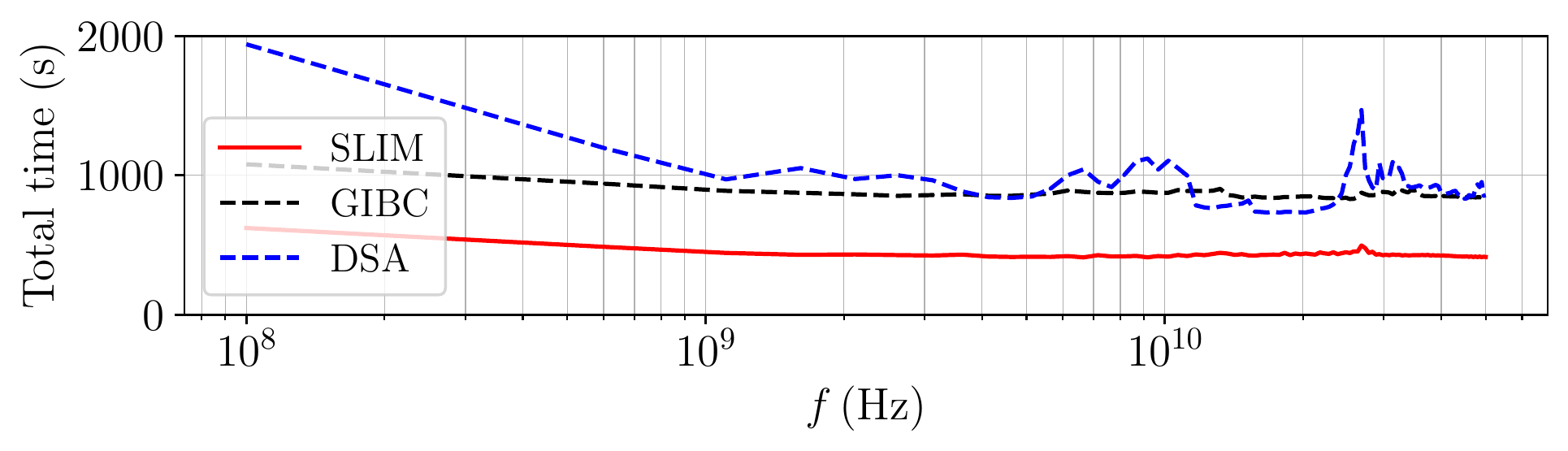}
  \caption{Performance comparison for the inductor array in \secref{sec:results:indarray}. \shortcolor{Top panel:} iterative solver convergence. \shortcolor{Bottom panel:} total simulation time.}\label{fig:indarray}
\end{figure}

Next, we consider a ${3 \times 4}$ array of copper inductor coils, excited via a Th\'evenin-equivalent port formulation~\cite{gope}, with ports defined as shown in \figref{fig:indarrayJ}.
\revcolor{Each inductor is a $4\times$ scaled version of the geometry considered in~\cite{EPEPS2017}.}
The inductors are embedded in the centre of a $50\,\mu$m dielectric layer with relative permittivity $2.1$.
Beneath the dielectric layer is a PEC-backed $50\,\mu$m layer of silicon with relative permittivity $11.9$ and conductivity $10\,$S/m.
The structure is meshed with $22,862$ triangles and $34,293$ edges.
Scattering parameters computed over a broad frequency range are shown in \figref{fig:indarrayS}, and are in excellent agreement with both the GIBC and DSA formulations.
\revcolor{The maximum error in $S_{11}$, $S_{12}$ and $S_{13}$ is $-75\,$dB, $-72\,$dB and $-74\,$dB, respectively, compared to the GIBC.}
The number of GMRES iterations required for convergence are reported in \shortcolor{the top panel of} \figref{fig:indarray}, which again highlights that the proposed formulation is as well conditioned as the GIBC, while the SSI formulation failed to converge within $800$ iterations.
Furthermore, as shown in \shortcolor{the bottom panel of} \figref{fig:indarray}, the SLIM formulation is faster than the GIBC since it does not require computing the double-layer potential operator in the outer problem.
In total, the SLIM formulation is over $2\times$ faster than both the GIBC and the DSA formulations.

\subsection{On-Chip Interconnect Network}\label{sec:results:interposer}

\begin{figure}[t]
  \centering
  \includegraphics[width=.90\linewidth,trim={0cm 0cm 0cm 0cm},clip=true]{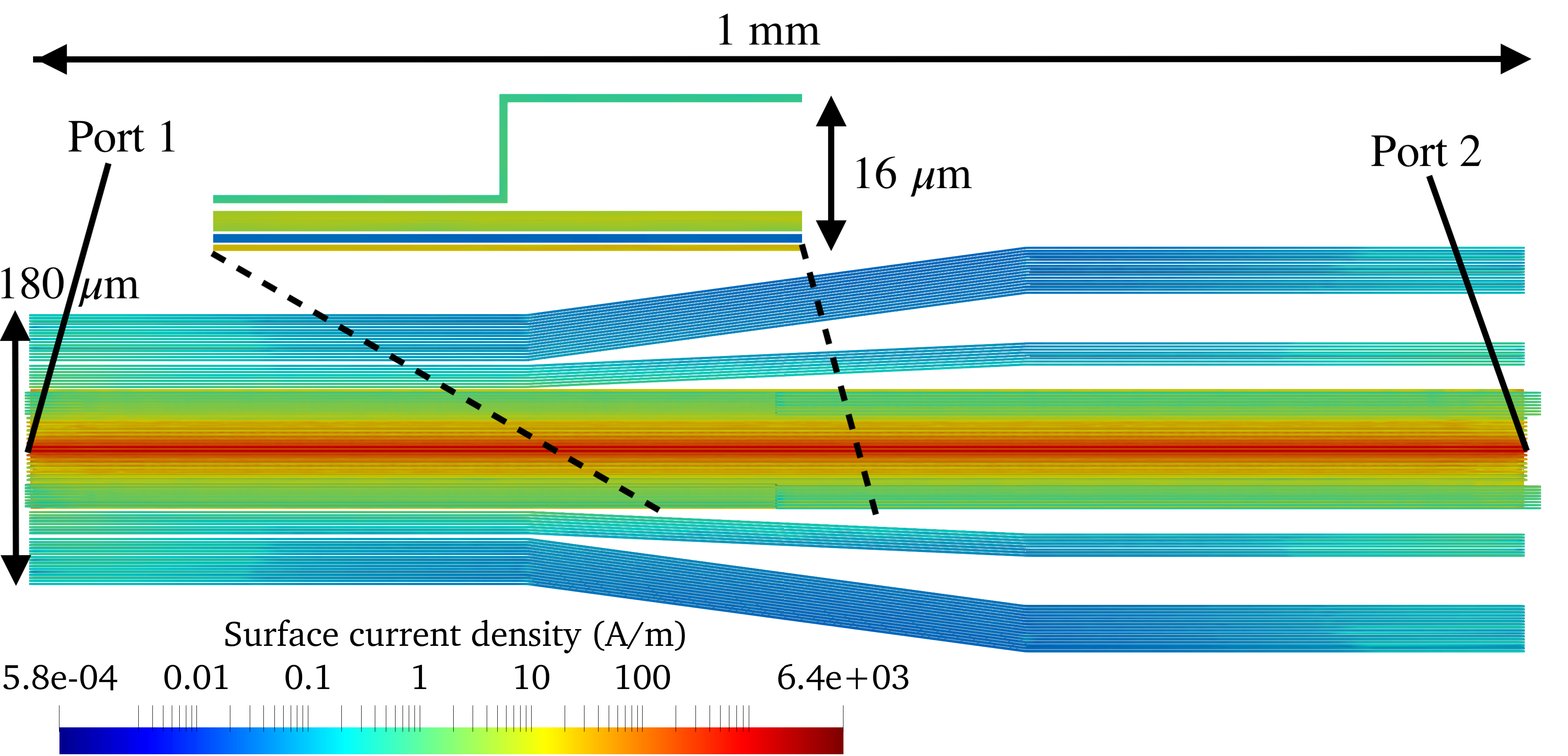}
  \caption{Geometry of the interconnect network in \secref{sec:results:interposer}, with the electric surface current density at $1\,$GHz.}\label{fig:J_interposer}
\end{figure}

\begin{figure}[th!]
  \centering
  \includegraphics[width=\linewidth]{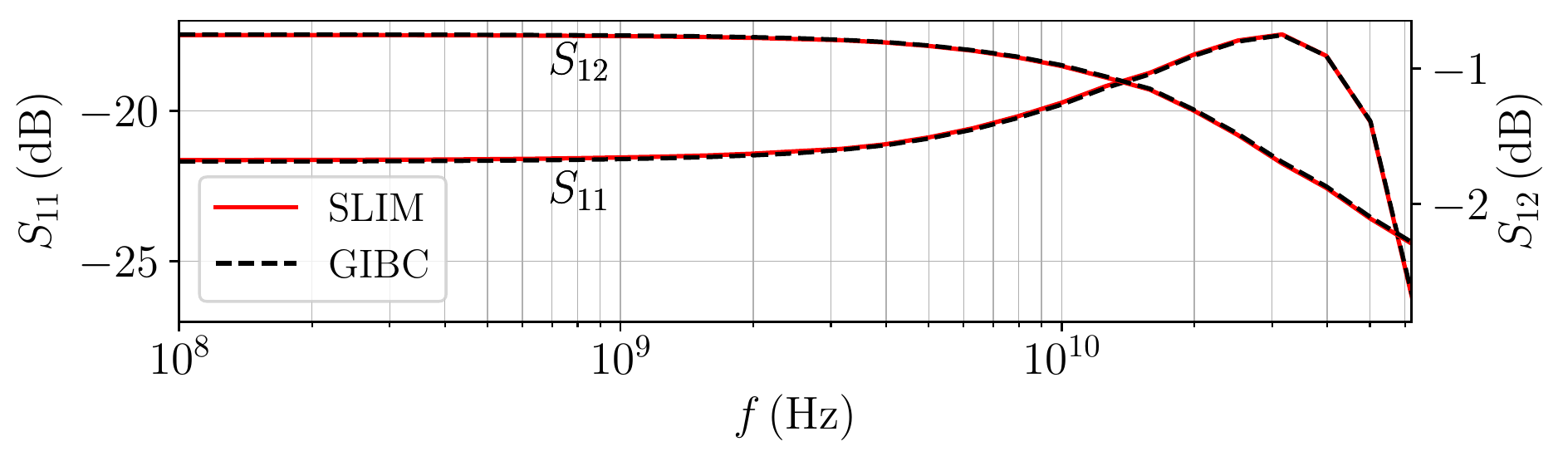}\\
  \includegraphics[width=\linewidth]{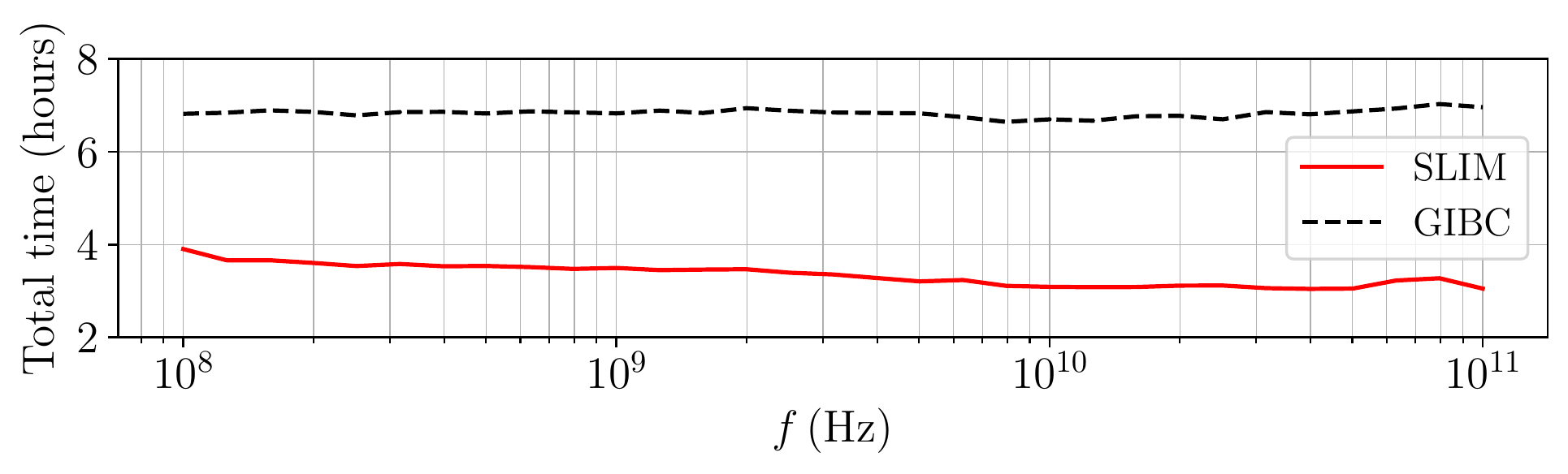}
  \caption{Comparison of the SLIM and GIBC methods for the interconnect network in \secref{sec:results:interposer}. \shortcolor{Top panel:} $S$ parameters. \shortcolor{Bottom panel:} total simulation time.}\label{fig:interposer}
\end{figure}

Finally, we consider an on-chip interconnect network consisting of 80 copper signal lines and a ground plane, embedded in a $27.5\,\mu$m thick dielectric layer with a relative permittivity of $4$ (courtesy of Dr. Rubaiyat Islam, Advanced Micro Devices).
Beneath the dielectric layer is a PEC-backed $500\,\mu$m layer of silicon with relative permittivity $11.9$ and conductivity of $10\,$S/m.
The structure is meshed with 156,820 triangles and 235,230 edges.
The geometry, port definitions and electric surface current density at $1\,$GHz are shown in \figref{fig:J_interposer}.
The $S$ parameters reported in \shortcolor{the top panel of} \figref{fig:interposer} confirm the accuracy of the SLIM formulation compared to the GIBC method.
\revcolor{The maximum error in $S_{11}$ and $S_{12}$ is $-37\,$dB and $-33\,$dB, respectively.}
The DSA and SSI formulations failed to converge within $800$ iterations, and therefore are not included in the results.
\shortcolor{The bottom panel of} \figref{fig:interposer} confirms that the proposed method is faster than the GIBC by a factor of over $2\times$, averaged over all frequency points.



\section{Conclusion}\label{sec:concl}
A novel single-source boundary element formulation is proposed for modeling lossy conductors in layered media.
Unlike existing techniques, the proposed formulation is well-conditioned without the need for dual basis functions, and avoids the double-layer potential in the exterior problem, making it more efficient when the background medium is modeled with the multilayer Green's function.
Numerical examples demonstrate that the proposed technique is more efficient than existing single- and dual-source techniques by a factor of at least $2\times$.

%
%

\ifCLASSOPTIONcaptionsoff
  \newpage
\fi



\bibliographystyle{ieeetr}
\bibliography{IEEEabrv,./bibliography}
\end{document}